# Generic full-vector angular spectrum method for calculating diffraction of arbitrary electromagnetic fields

CHENGDA SONG, JING HE, AND GUANGHUI YUAN*

*Department of Optics and Optical Engineering, School of Physical Sciences, University of Science and Technology of China, Hefei, Anhui 230026, China*
*ghyuan@ustc.edu.cn

**Abstract:** Numerous vector angular spectrum methods have been presented to model the vectorial nature of diffractive electromagnetic field, facilitating optical field engineering in polarization-related and high numerical aperture systems. However, balancing accuracy and efficiency in state-of-the-art vector methods is challenging, especially with not well-defined incident fields. Here, we propose a full-vector angular spectrum method for accurate, efficient, robust diffraction computation, allowing truly arbitrary incidence by precisely modeling the projection rule among Cartesian polarization components. We address a prior oversight, that the longitudinal electric field's projection onto the diffracted field was insufficiently considered. Notably, our method inherently handles reflection and transmission at dielectric interfaces, which can be viewed as k-space filters. For rotationally symmetric system, it achieves unprecedented computation times of a few seconds, speeding up optical design via faster input-output mapping in optimization algorithms.

## 1. INTRODUCTION

Accurate modeling of free-space diffraction of arbitrary electromagnetic field is a fundamental and key issue in modern optics, especially at micro/nano scales. Scalar diffraction theories fail to capture the rich vectorial nature of electromagnetic fields, a critical limitation in high-precision applications, such as tight focusing system [1–6], light field engineering in topological optics and structured light fields [7–15], spin angular momentum [16–18], polarization and orbital angular momentum multiplexing in holography and encryption [19–22]. These demand consideration of vectorial effects — polarization coupling, longitudinal field components, and angular momentum transfer — which challenge existing theories.

Traditional vector diffraction theories split into two branches that view wavefronts as different superpositions of point sources or plane waves. The Dyad Green's Function (DGF) method in real space excels with point sources but is computationally costly for arbitrary wavefronts [23,24]. The angular spectrum method (ASM) in k-space, more common due to its superior computational performance and clear physical meaning, decomposes wavefronts into plane waves propagating in three-dimensional directions via Fast Fourier Transform (FFT) [25–27]. Moreover, for high-NA focusing, J. Lin *et al*. rewrote the Debye-Wolf integral as a 3D convolution and thus developed a 3D-ASM to perform fast and accurate calculations of the 3D vectorial field distribution [28,29], which was then extended to treat Rayleigh-Sommerfeld diffraction formula without approximations [30]. Yet, typical ASMs [25–27] assume that the electric field components propagate independently, neglecting the inherent cross-polarization coupling. This pseudo-vectorial approach applies scalar methods to each component despite vectorial plane waves being transversal with only two degrees of freedom of polarization. P. Török *et al*. [31] and S. W. Hell *et al*. [32] showed its flaws at dielectric interfaces in confocal systems, favoring DGF for polarization interactions.

To address these limitations, researchers have refined ASM. Mansuripur [33] added a 3×2 transition matrix for cross-polarization conversion, mapping incident transverse polarization (*x*-, *y*-terms) to diffracted polarization (*x*-, *y*-, *z*-terms). Effective and efficient for certain vector diffraction problems [34], it mismatches Helmholtz equation and free-divergence conditions ( $\nabla \cdot \vec{E} = 0$ ). Davis' vector potential [35,36] and Hertz vector [37,38] approximation are then introduced and advanced studies on Gaussian waves. Rosario *et al*. [39] proposed an integral ASM, decomposing incident field into s- and p-polarized terms. These methods improve accuracy over common ASM but falter with high numerical aperture (NA) cases or increase computational demands without a closed FFT form, often ignoring longitudinal polarization (z-term) and dielectric interface effects.

In this work, we present a generalized full-vector ASM (VASM) framework for free-space vectorial diffraction of arbitrary electromagnetic fields, offering unmatched computational accuracy and efficiency. Combining a projection matrix with s-p wave decomposition, we derived an explicit, complete and symmetrical 3×3 FFT-based formula. Polarization interactions and conversions in the projection matrix have endowed our VASM with strong robustness. Especially, with one input polarization component (others zero), our method yields all the six diffracted field profiles (three electric, three magnetic). It aligns well with full-wave simulations like finite-difference time-domain (FDTD), reducing computation times from hours to minutes, or even seconds with certain symmetry arrangement. Our s-p decomposition naturally models planar interface reflection and transmission, integrating Fresnel coefficients as k-space modulation in the FFT framework.

The paper is organized as follows. Section 2 derives our generic 2D FFT-based VASM formulae, introduces a simplified 1D-integral form for rotationally symmetric cases, and addresses planar interface reflection and transmission issues. Section 3 tests accuracy of our VASM under high-NA conditions, comparing our VASM, common ASM and FDTD. Section 4 explores rotationally symmetric cases, overcoming the limitations in 2D-FFT algorithm, and implements a numerical experiment to define the boundaries between scalar and vector theories.

## 2. FULL-VECTOR ASM

Conventional scalar diffraction theory fails to model high NA focusing and vectorial features. We developed a rigorous VASM method, decomposing fields into s- and p-polarizations. By projecting between Cartesian and intrinsic bases, we derived a complete k-space representation with a compact matrix for vector fields propagation. Key results include a generalized FFT-based VASM and handling of planar dielectric interfaces with Fresnel coefficients, laying the foundation for efficient vectorial focusing and spectral filtering.

*2.1 Basic theory of vector angular spectrum method*

Before introducing vector cases, a brief review of scalar ASM is considered instructive. Any scalar incident field can be decomposed into a series of plane-waves. Assuming an incident scalar field $E(x_o, y_o, 0)$ propagates along $z$, the scalar ASM method gives the diffracted field $E(x_i, y_i, z_f)$ after distance $z_f$:

$$E(x_i, y_i, z) = \mathcal{F}^{-1}\left\{\mathcal{F}\{E(x_o, y_o, 0)\}e^{i\sqrt{k^2-(k_x^2+k_y^2)}z}\right\} \qquad (1)$$

Here $\mathcal{F}\{\cdot\}$ and $\mathcal{F}^{-1}\{\cdot\}$ denotes FFT and inverse FFT (iFFT), with $(x,y)$ and $(k_x, k_y)$ as real and k-space coordinates respectively. Wavenumber $k \equiv nk_0$, where $n$ is the medium's refractive index and $k_0$ the vacuum wavenumber. The spectrum $\mathcal{E}(k_x, k_y) \equiv \mathcal{F}\{E(x_o, y_o, 0)\}$ provides each constituent plane-wave's amplitude and phase, termed the angular spectrum. For propagating waves (neglecting evanescent ones), the spectrum $\mathcal{E}(k_x, k_y)$ is band-limited in k-space: $\sqrt{k_x^2 + k_y^2} < k$.

The aforementioned plane-wave decomposition can be directly extended to vectorial cases. By applying scalar ASM to each cartesian component, common treatments in textbooks [26,27] yield the commonly used vector ASM:

$$\mathbf{E}(x_i, y_i, z) = \mathcal{F}^{-1}\left\{\mathbb{I} \cdot \mathcal{F}\{\mathbf{E}(x_o, y_o, 0)\} e^{i\sqrt{k^2 - (k_x^2 + k_y^2)}z}\right\} \quad (2)$$

Here $\mathbb{I}$ is a 3×3 identity matrix, $\mathbf{E}(x_o, y_o, z) = (E_x, E_y, E_z)^T$ in Cartesian basis, 'T' denotes transpose.

Although the common ASM (Eq. (2)) rigorously satisfies the vector Helmholtz Equation, it must additionally fulfill the free-divergence condition $\nabla \cdot \mathbf{E} = 0$ to fully comply with Maxwell's Equations. This imposes strong constraints on the incident field $\mathbf{E}$, which is often not well-defined. For example, in many planar metalens design cases, a linearly polarized incident beam $\mathbf{E} = (E_x, 0, 0)^T$ with a converging spherical wavefront is typically assumed, which clearly violates $\nabla \cdot \mathbf{E} = 0$.

Another notable approach proposed by B. Richards and E. Wolf [1] introduces an alternative decomposition method. By combining diffraction with the simple law of refraction, the converging spherical wavefront with a focal length $f$ can be decomposed from cylindrical coordinate system ($\rho, \phi, z$; incident beam) into spherical coordinate system ($r, \phi, \theta$; focusing beam after refraction). The tightly-focused field can then be calculated by the well-known Richards-Wolf integral:

$$\mathbf{E}(x_i, y_i, z) = -\frac{ife^{-ikf}}{2\pi}\mathcal{F}^{-1}\left\{\begin{pmatrix} \frac{k_x^2}{k_t^2}\frac{k_z}{k} + \frac{k_y^2}{k_t^2} & \frac{k_x k_y}{k_t^2}(\frac{k_z - k}{k}) & -\frac{k_x}{k} \\ \frac{k_x k_y}{k_t^2}(\frac{k_z - k}{k}) & \frac{k_y^2}{k_t^2}\frac{k_z}{k} + \frac{k_x^2}{k_t^2} & -\frac{k_y}{k} \\ \frac{k_x}{k} & \frac{k_y}{k} & \frac{k_z}{k} \end{pmatrix} \cdot \frac{\mathbf{E}_{\text{inc}}(k_x, k_y)}{\sqrt{kk_z}}e^{ik_z z}\right\} \quad (3)$$

where transverse wavenumber $k_t = \sqrt{k_x^2 + k_y^2}$ and longitudinal wavenumber $k_z = \sqrt{k^2 - k_t^2}$. The spectrum $\mathbf{E}_{\text{inc}}(k_x, k_y)$ in tight focusing cases is exactly the incident field $\mathbf{E}(x, y, 0)$, with the relation $k_x = x_o k/f, k_y = y_o k/f$. Richards-Wolf integral successfully models the vectorial interactions under high-NA focusing system. However, Eq. (3) remains restricted to bulk, spherical aberration-free, and paraxial focusing system [2–6], making it unsuitable for modeling planar diffraction meta-elements.

Following Richards and Wolf's approach, the key challenge lies in determining a set of appropriate decomposition basis vectors $\vec{\alpha}$, where the transverse condition $\nabla \cdot \mathbf{E} = 0$ is inherently satisfied. We choose a triad of mutually perpendicular right-handed system of unit vectors: $\vec{\alpha} = \boldsymbol{\sigma}, \boldsymbol{s}, \boldsymbol{p}$ with $\boldsymbol{\sigma} = (k_x, k_y, k_z)/k$, $\mathbf{s} = (k_y, -k_x, 0)/k_t$, $\mathbf{p} = (k_z k_x, k_z k_y, -k_t^2)/(kk_t)$. Here $\boldsymbol{\sigma} = \boldsymbol{k}/|\boldsymbol{k}|$ is the normalized wavevector, $\mathbf{s}$ and $\mathbf{p}$ correspond to the s- and p- unit vector respectively. Thus, the angular spectra in Cartesian coordinates system can be decomposed into $\vec{\alpha} = \boldsymbol{\sigma}, \boldsymbol{s}, \boldsymbol{p}$ as:

$$\mathcal{E}(k_x, k_y, z) = \mathcal{E}_x \mathbf{e}_x + \mathcal{E}_y \mathbf{e}_y + \mathcal{E}_z \mathbf{e}_z = (\mathcal{E} \cdot \mathbf{s})\mathbf{s} + (\mathcal{E} \cdot \mathbf{p})\mathbf{p} \equiv \mathcal{E}_s \mathbf{s} + \mathcal{E}_p \mathbf{p} \quad (4)$$

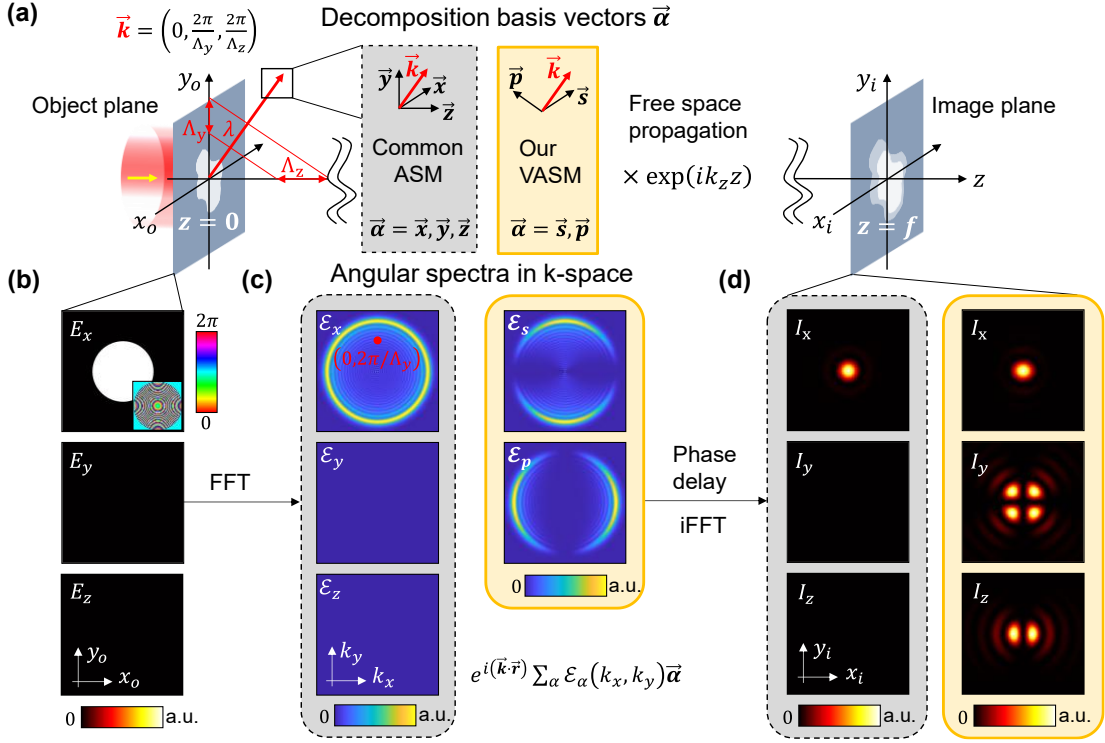

Fig. 1. (a) Schematic of common ASM and our VASM, framed by dashed grey box and solid yellow box, respectively. (b) Planar metalens with a converging phase. (c) Angular spectra $\mathcal{E}_\alpha$ of common ASM and our VASM. (d) Intensity of vector field at focal plane, calculated by common ASM and our VASM. All profiles are scaled to the same amplitude.

It is worth noting that the spectra $\mathcal{E}_s$ and $\mathcal{E}_p$ require additional superposition to reconstruct Cartesian field components. The full process — decomposition, propagation, superposition — can be articulated as an explicit FFT-based VASM formula:

$$\mathbf{E}(x_i, y_i, z) = \mathcal{F}^{-1}\left\{\begin{pmatrix} 1-\sigma_x^2 & -\sigma_x\sigma_y & -\sigma_x\sigma_z \\ -\sigma_x\sigma_y & 1-\sigma_y^2 & -\sigma_y\sigma_z \\ -\sigma_x\sigma_z & -\sigma_y\sigma_z & 1-\sigma_z^2 \end{pmatrix} \cdot \mathcal{F}\{\mathbf{E}(x_o, y_o, 0)\}e^{ik_z z}\right\} \quad (5)$$

Here, $\boldsymbol{\sigma} \equiv (\sigma_x, \sigma_y, \sigma_z)$ corresponds to the normalized wavevector. For completeness, diffracted magnetic fields $\mathbf{H}$ can also be derived from Maxwell's Equations $\mathcal{H} = \mathcal{F}\{\mathbf{H}\} = Z_{\mu\varepsilon}^{-1}(\boldsymbol{\sigma} \times \boldsymbol{\mathcal{E}})$:

$$\mathbf{H}(x_i, y_i, z) = Z_{\mu\varepsilon}^{-1}\mathcal{F}^{-1}\left\{\begin{pmatrix} 0 & -\sigma_z & \sigma_y \\ \sigma_z & 0 & -\sigma_x \\ -\sigma_y & \sigma_x & 0 \end{pmatrix} \cdot \mathcal{F}\{\mathbf{E}(x_o, y_o, 0)\}e^{ik_z z}\right\} \quad (6)$$

$Z_{\mu\varepsilon} = \sqrt{\mu_0\mu/\varepsilon_0\varepsilon}$ is the wave impedance of the medium, with $\mu_0$ and $\varepsilon_0$ being the free-space permeability and permittivity, $\mu$ and $\varepsilon$ being medium values. Equations (5) and (6) form our full-field (electric fields and consistent magnetic fields), FFT-based VASM formulae for calculation of diffracted electromagnetic fields of arbitrary incidence.

Figure 1 illustrates the schematic of the calculation models, highlighting the key difference between common ASM and our VASM. A real-space plane wave with a space period $\boldsymbol{\Lambda} = (\infty, \Lambda_y, \Lambda_z)$ can be denoted as $\mathcal{E}(0, 2\pi/\Lambda_y)e^{i2\pi(y/\Lambda_y + z/\Lambda_z)}$, corresponding to a k-space coordinate $(k_x, k_y) = 2\pi(0, 1/\Lambda_y)$.

The decomposition basis vectors of common ASM ($\mathbf{x}, \mathbf{y}, \mathbf{z}$) and our VASM ($\boldsymbol{\sigma}, \mathbf{s}, \mathbf{p}$) are framed in the dashed grey box in Fig. 1(a).

By reformulating the divergence-free condition in k-space as $\mathbf{k} \cdot \boldsymbol{\mathcal{E}} = \mathbf{0}$, it becomes evident that for a well-defined vector plane wave $\boldsymbol{\mathcal{E}}(k_x, k_y)$, the possible polarization directions are restricted to the plane perpendicular to $\mathbf{k}$. Consequently, the polarization only has two degrees of freedom, rendering the Cartesian basis vectors $\vec{\alpha} = \mathbf{x}, \mathbf{y}, \mathbf{z}$ used in common ASM unsuitable. In contrast, while the basis vectors $\vec{\alpha} = \boldsymbol{\sigma}, \mathbf{s}, \mathbf{p}$ shown in the solid yellow box exhibit exactly two degrees of freedom. This demonstrates how the constraint on incident vector fields is transformed into a constraint on basis vectors. When properly selected, these basis vectors can accommodate truly arbitrary incidence while ensuring robustness.

Figure 1(b) depicts a horizontally polarized vector field with a converging phase $\mathbf{E} = \left(\text{circ}(r_o/R) \exp ik\left(f - \sqrt{f^2 + r_o^2}\right), 0, 0\right)^T$, which clearly violates the divergence-free condition $\nabla \cdot \mathbf{E} = 0$. In common ASM, the diffracted vector field manifests as a horizontally polarized Airy pattern due to the absence of other polarization components, $E_y$ and $E_z$ (framed in the dashed grey box in Figs. 1 (c) & 1(d)). While one might attempt to reconstruct appropriate vector spectra using the condition $\mathcal{E}_z = -k_x \mathcal{E}_x / k_z$, this approach requires prior knowledge and offers no practical advantage. Furthermore, this method still fails to capture the vectorial distortions in the diffracted $E_x(x_i, y_i, z)$, which maintains its Airy pattern characteristics.

In contrast, our VASM successfully retrieves all vector features without introducing any approximations, as shown in the solid yellow box in Fig. 1(d). All profiles are scaled to the same amplitude for simplicity. The focused intensity distribution shows excellent agreement with Richards-Wolf theory exhibiting x-axis elongation in $I_x$, a four-lobe pattern in $I_y$ and a two-lobe pattern in $I_z$, all of which have been experimentally verified. Importantly, our VASM works without approximations, enabling direct calculation of diffracted field for arbitrary incident beam $\mathbf{E}(x_o, y_o, 0)$. This capability makes it particularly suitable for analyzing the diffraction in high-NA planar meta-devices.

Furthermore, our method explicitly incorporates for the often-ignored z-polarization component. The inherent symmetry of the projection matrix indicates that no single polarization component dominates, and all components must be considered. Neglecting $E_z(x, y, 0)$ at the incident plane would lead to discrepancies in the detected $E_x(x_i, y_i, z)$ and $E_y(x_i, y_i, z)$ at the receiving plane, as the matrix's third column in Eq. (5) is non-zero, representing p-wave to $E_z$ conversion. These discrepancies become particularly significant when $E_z$ becomes comparable to or exceeds $E_x$ and $E_y$ — a crucial advancement that prior studies have not adequately addressed.

It should be emphasized that common ASM represents not an incorrect but rather a simplified approach. Any incident field fulfilling $\nabla \cdot \mathbf{E} = 0$ would lead to degeneration, for instance:

$$\mathcal{E}'_x = (1 - \sigma_x^2)\mathcal{E}_x - \sigma_x \sigma_y \mathcal{E}_y - \sigma_x \sigma_z \mathcal{E}_z = \mathcal{E}_x - \frac{\sigma_x}{k}(\mathbf{k} \cdot \boldsymbol{\mathcal{E}}) = \mathcal{E}_x \qquad (7)$$

Here $\mathcal{E}'_x$ is the post-superposition *x*-component of the spectrum, corresponding to the first row of Eq. (5). Equation (7) clearly reveals that the Cartesian vector spectrum appears to propagate independently, but only under rigorous divergence-free conditions. Physically, conversions among the components cancel each other out delicately. Under general conditions such subtlety easily breaks, and we must turn to the original interaction matrix in Eq. (5) for solutions, a situation we often encounter when calculating electromagnetic field diffraction problems near a dielectric interface.

## 2.2 Diffraction of rotationally symmetric planar masks

For planar diffraction masks with rotational symmetry, Eq. (5) can be simplified to an integral form. Assuming a linear polarized (LP) electrical field $\mathbf{E}(x_o, y_o, 0) = E_0(x_o, y_o, 0)\mathbf{e}_x$ as an example, by inserting it into Eq. (5) under cylindrical coordinates, we get:

$$\mathbf{E}(r_i, \theta_i, z) = \frac{1}{(2\pi)^2 k^2} \int_0^k dk_t \int_0^{2\pi} d\varphi \begin{pmatrix} k^2 - k_t^2 \cos^2\varphi \\ -\cos\varphi \sin\varphi\, k_t^2 \\ -\cos\varphi\, k_t k_z \end{pmatrix} \tilde{E}_0(k_t, \varphi) e^{ik_z z} e^{ik_t r_i \cos(\varphi - \theta_i)} k_t \quad (8)$$

where $\tilde{E}_0(k_t, \varphi) = \int_0^R E_0(r_o, \theta_o, 0) e^{-ik_t r_o \cos(\varphi - \theta_o)} r_o dr_o d\theta$, $k_z = \sqrt{k^2 - k_t^2}$ and $(r_i, \theta_i)$ are the polar coordinates in the receiving plane. Obviously when $E_0(r_o, \theta_o, 0)$ is rotationally symmetric, $\tilde{E}_0(k_t, \varphi)$ is independent on azimuthal angle $\varphi$, Eq. (8) becomes:

$$\mathbf{E}(r_i, \theta_i, z) = \frac{1}{4\pi k^2} \int_0^k dk_t \begin{bmatrix} 2J_0(k_t r_i)k^2 - \left(J_0(k_t r_i) - J_2(k_t r_i) \cos 2\theta_i\right)k_t^2 \\ 2\cos\theta_i \sin\theta_i J_2(k_t r_i) k_t^2 \\ -2iJ_1(k_t r_i) \cos\theta_i\, k_t k_z \end{bmatrix} \widetilde{E}_0(k_t) e^{ik_z z} k_t \quad (9)$$

$J_n(\cdot)$ is Bessel function of the first kind of order $n$. The $\theta_i$-dependent terms in Eq. (9) cause elongation and lobe-like patterns in $E_x$, $E_y$ and $E_z$ shown in Fig. 1(d). If $E_x$ is dominant, the focus distortion behaves as an elongation in the polarization direction. Tighter (higher NA) focusing results in more elliptical focal spots.

A feasible approach to conserve the circularity of the focus is to illuminate the mask with a circularly polarized (CP) beam. We derive the diffracted field of a left-handed circular polarized (LCP) beam $E_0(x_o, y_o, 0)\mathbf{e}_x + iE_0(x_o, y_o, 0)\mathbf{e}_y$:

$$\mathbf{E}(r_i, \theta_i, z) = \frac{1}{4\pi k^2} \int_0^k dk_t \begin{bmatrix} 2J_0(k_t r_i)k^2 - \left(J_0(k_t r_i) - J_2(k_t r_i)e^{i2\theta_i}\right)k_t^2 \\ 2iJ_0(k_t r_i)k^2 - i\left(J_0(k_t r_i) + J_2(k_t r_i)e^{i2\theta_i}\right)k_t^2 \\ -2iJ_1(k_t r_i)e^{i\theta_i} k_t k_z \end{bmatrix} \widetilde{E}_0(k_t) e^{ik_z z} k_t \quad (10)$$

Using the following abbreviations for integrals involved:

$$\begin{cases} F_0(r_i, z) = \frac{1}{4\pi k^2} \int_0^k dk_t J_0(k_t r_i) k^2 \widetilde{E}_0(k_t) e^{ik_z z} k_t \\ F_0'(r_i, z) = \frac{1}{4\pi k^2} \int_0^k dk_t J_0(k_t r_i) k_t^2 \widetilde{E}_0(k_t) e^{ik_z z} k_t \\ F_1(r_i, z) = \frac{1}{4\pi k^2} \int_0^k dk_t J_1(k_t r_i) k_t k_z \widetilde{E}_0(k_t) e^{ik_z z} k_t \\ F_2(r_i, z) = \frac{1}{4\pi k^2} \int_0^k dk_t J_2(k_t r_i) k_t^2 \widetilde{E}_0(k_t) e^{ik_z z} k_t \end{cases} \quad (11)$$

Eq. (11) can be expressed as a variable-separated form:

$$\mathbf{E}(r_i, \theta_i, z) = \begin{bmatrix} 2F_0(r_i, z) - F_0'(r_i, z) + F_2(r_i, z)e^{i2\theta_i} \\ 2iF_0(r_i, z) - iF_0'(r_i, z) - iF_2(r_i, z)e^{i2\theta_i} \\ -2iF_1(r_i, z)e^{i\theta_i} \end{bmatrix} \quad (12)$$

According to Eq. (12), the diffracted field of LCP incidence contains both LCP and right-handed circular polarized (RCP) components. The total intensity of detectable transverse fields, $E_x$ and $E_y$, exhibits rotational symmetry at any propagation distance $z$:

$$I_{\text{trans}} = I_x + I_y \propto (4|F_0|^2 + |F_0'|^2 - 4\Re\{F_0 F_0'^*\} + |F_2|^2) \quad (13)$$

Another interesting observation is that the total intensity predicted in scalar theory is exactly the first term $4|F_0|^2$. When the incident field is circularly polarized, the transverse diffraction patterns can be well-estimated by scalar theory.

## 2.3 Reflection and transmission of planar interface

Optics often involves planar surfaces — e.g., immersion lenses or optical tweezers. Planar diffraction meta-lenses need extra dielectric layers to maintain resonance in liquids. Fresnel coefficients, expressed via incidence angle or wavevectors, describe refraction and transmission:

$$\begin{cases} r^s(k_x,k_y) = \frac{1-\frac{\mu_1 k_{z2}}{\mu_2 k_{z1}}}{1+\frac{\mu_1 k_{z2}}{\mu_2 k_{z1}}}, & r^p(k_x,k_y) = \frac{1-\frac{k_{z2}\epsilon_1}{k_{z1}\epsilon_2}}{1+\frac{k_{z2}\epsilon_1}{k_{z1}\epsilon_2}} \\ t^s(k_x,k_y) = \frac{2}{\left(1+\frac{\mu_1 k_{z2}}{\mu_2 k_{z1}}\right)}, & t^p(k_x,k_y) = \sqrt{\frac{\mu_2 \epsilon_1}{\mu_1 \epsilon_2}} \frac{2}{\left(1+\frac{k_{z2}\epsilon_1}{k_{z1}\epsilon_2}\right)} \end{cases} \quad (14)$$

Transverse wavevectors $(k_x, k_y)$ are conserved according to the boundary conditions, and longitudinal wavenumbers $k_{zi} = \sqrt{(n_i k_0)^2 - k_x^2 - k_y^2}$ varies with medium constants $\mu_i$ and $\varepsilon_i$. Projection-based VASM method is inherently prominent because we can handle the interface effects by simply multiplying the projection matrix by a few $(k_x, k_y)$-related factors in k-space:

$$\widetilde{\mathcal{E}}_t = (t^s \mathbf{ss} + t^p \mathbf{pp}) \cdot \widetilde{\mathcal{E}}, \quad \widetilde{\mathcal{E}}_r = (r^s \mathbf{ss} + r^p \mathbf{pp}) \cdot \widetilde{\mathcal{E}} \quad (15)$$

where $\widetilde{\mathcal{E}}_t$ and $\widetilde{\mathcal{E}}_r$ are the spectra of transmissive and reflected fields respectively.

Since $r^s \neq r^p$ and $t^s \neq t^p$, the updated matrix in Eq. (5) never degenerates into identity matrix under any circumstances as mentioned above, highlighting the advantage of the VASM. Interface acts as angle-dependent modulator or k-space filter, decoupling transverse field intensity profile and polarization with a thin dielectric film.

Equation (15) quantifies substrate effects on diffracted fields, which is critical for designing cascaded meta-lens. In low-NA and paraxial approximation, the influence of substrate is negligible due to small $k_t$. It's reasonable to neglect the non-diagonal elements in the projection matrix, and the coefficients in Eq. (14) are constants. In high-NA, total internal reflection (TIR) and multiple reflections should be considered. The spectrum passing through a substrate with thickness of $d$ can be written in terms of direct transmission and that after multiple reflections:

$$\widetilde{\mathcal{E}}'^{(i)} = \widetilde{\mathcal{E}}^{(i)} e^{ik_z d} t_1^{(i)} t_2^{(i)} \left( \sum_{n=0}^{N-1} \left( e^{i2k_z d} r_2^{(i)} r_1^{(i)} \right)^n \right) = \widetilde{\mathcal{E}}^{(i)} e^{ik_z d} t_1^{(i)} t_2^{(i)} \left( \frac{1-e^{i2Nk_z d} r_2^{(i)} r_1^{(i)}}{1-e^{i2k_z d} r_2^{(i)} r_1^{(i)}} \right) \quad (16)$$

$\widetilde{\mathcal{E}}^{(i)}$ is the incident spectrum of either s- or p-polarization. $t_1^{(i)}$ and $r_1^{(i)}$ are the Fresnel coefficients after the medium/substrate interface, while $t_2^{(i)}$ and $r_2^{(i)}$ represent those before the medium/substrate interface (see the model in Section 3 for details). Neglecting higher order terms will lead to inaccuracies such as focus shift or size of focal spot. For multiple layers, transfer matrix method and our VASM can be jointly used to calculate the final transmissive or reflective fields.

Our VASM offers efficiency and simplicity for high-NA focusing and interface effects in k-space. Using the s-p basis, it models vectorial behaviors beyond scalar methods. Next, we validate this through numerical simulations and demonstrate its design utility.

## 3. VERIFICAITON CASES

In this section we quantitatively validate our VASM method against FDTD simulations, showcasing its ability to model high-NA focusing with dielectric interfaces. We also demonstrate its flexibility in analyzing transmission and reflection at dielectric interfaces, which is vital for designing immersion lenses, anti-reflection coatings and interface-dependent optical components. Benchmarking accuracy affirms our VASM as a powerful tool for photonics applications with vectorial behaviors.

### 3.1 Direct propagation

For an example of high-NA planar meta-devices, we designed a binary amplitude mask with concentric rings to test our method against the ground truth calculated by FDTD. Such planar diffraction optical elements disobey the approximation made in Richards-Wolf integral, and one must turn to angular spectrum method instead. As shown in Fig. 2(a), the mask has a radius of 10 μm and is designed to focus incident plane-wave ($\lambda = 633$ nm) to $z = 10$ μm. The red arrows indicate the polarization of the incident field, here chosen as $\vec{x}$. The mask has 25 concentric rings, each has a uniform width of 0.4 μm, either transparent or opaque, optimized with particle swarm optimization algorithm. To maximize vectorial behaviors, the mask is submerged in water (n=1.33) to reach a higher NA of 0.94.

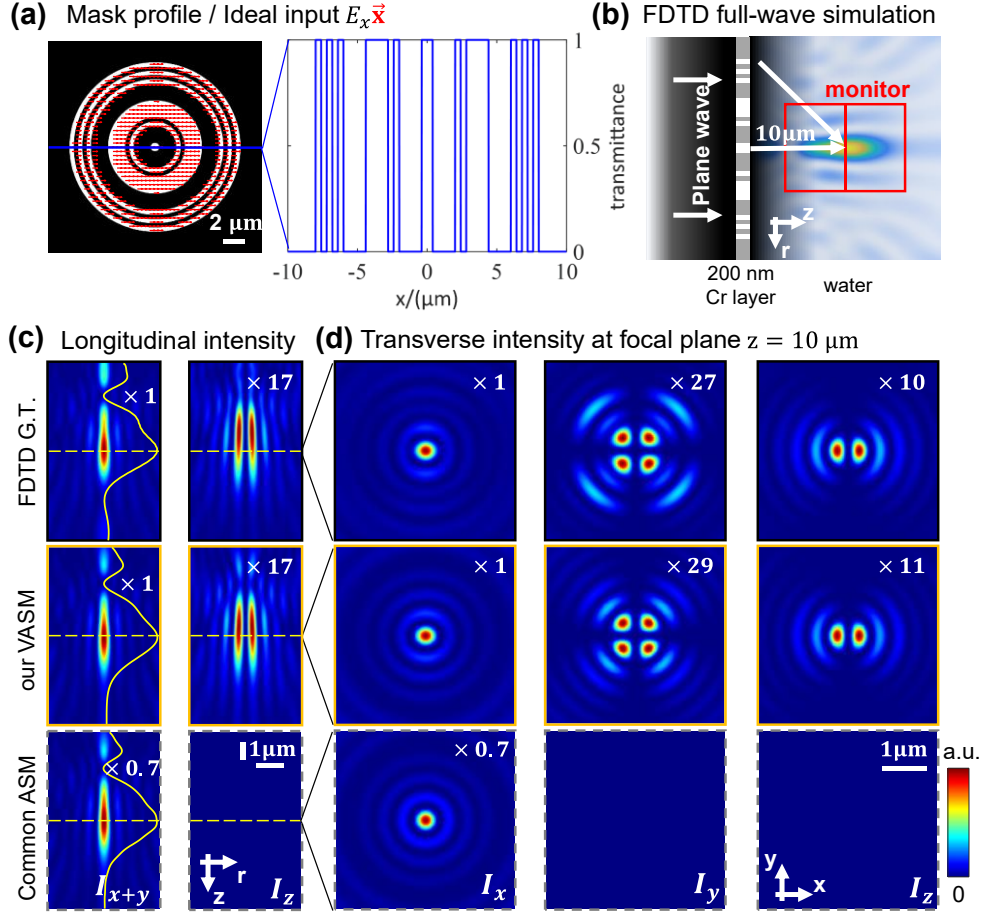

Fig. 2. Simulation results of a designed mask. (a) Binary amplitude mask with 25 concentric 0.4 μm wide rings. Incident wavelength: 633 nm, x-polarization incidence. (b) Schematic diagram of FDTD. (c) Longitudinal $|\mathbf{E}|^2$ profiles calculated by FDTD, our VASM and common ASM. 'G.T.' is the abbreviation to 'Ground Truth'. (d) Transversal $|\mathbf{E}|^2$ profiles calculated by FDTD, our VASM and common ASM. All profiles are scaled to the same amplitude with scaling factors in top-right corners.

We first demonstrate the robustness of VASM by restricting the simulation inputs. Both our VASM and common ASM employ an ideal mask (Fig. 2(a)), which assumes perfect illumination conditions while neglecting material response and boundary effects. In contrast, the FDTD simulation (Fig. 2(b)) incorporates realistic experimental conditions, including a 200-nm-thick chromium film with physically etched mask patterns. This systematic comparison enables rigorous assessment of our method's performance across both idealized and practical experimental scenarios.

We anticipate that our VASM is capable for retrieving all three components of vector fields while the input is not well-defined. We compared the up-mentioned 3 different models by respectively calculating the electric field intensity profiles on longitudinal r-z plane (5 μm span around focus) and transversal focal planes. The results are given in Figs. 2(c) and 2(d) respectively, with FDTD's results in the first row, and our VASM's and common ASM's in the second and the last column, which are framed by solid yellow boxes and dashed grey boxes respectively. For simplicity, all profiles are scaled to the same amplitude with scaling factors in top-right corners. In addition, the central intensity distribution $I(0,0,z)$ is also inserted into Fig. 2(c), plotted as solid light-yellow curves, while the focal plane is indicated by dashed light-yellow lines.

The results show quantitative agreement between FDTD and our VASM. The transversal intensity profiles show excellent correlation, particularly in the critical central region. Slight difference along the z-axis in Fig. 2(c) is explainable due to non-ideal illumination and realistic material response of chromium layer. This reflects the robustness of our method since we did not add any prior knowledge about the possible response to our VASM model.

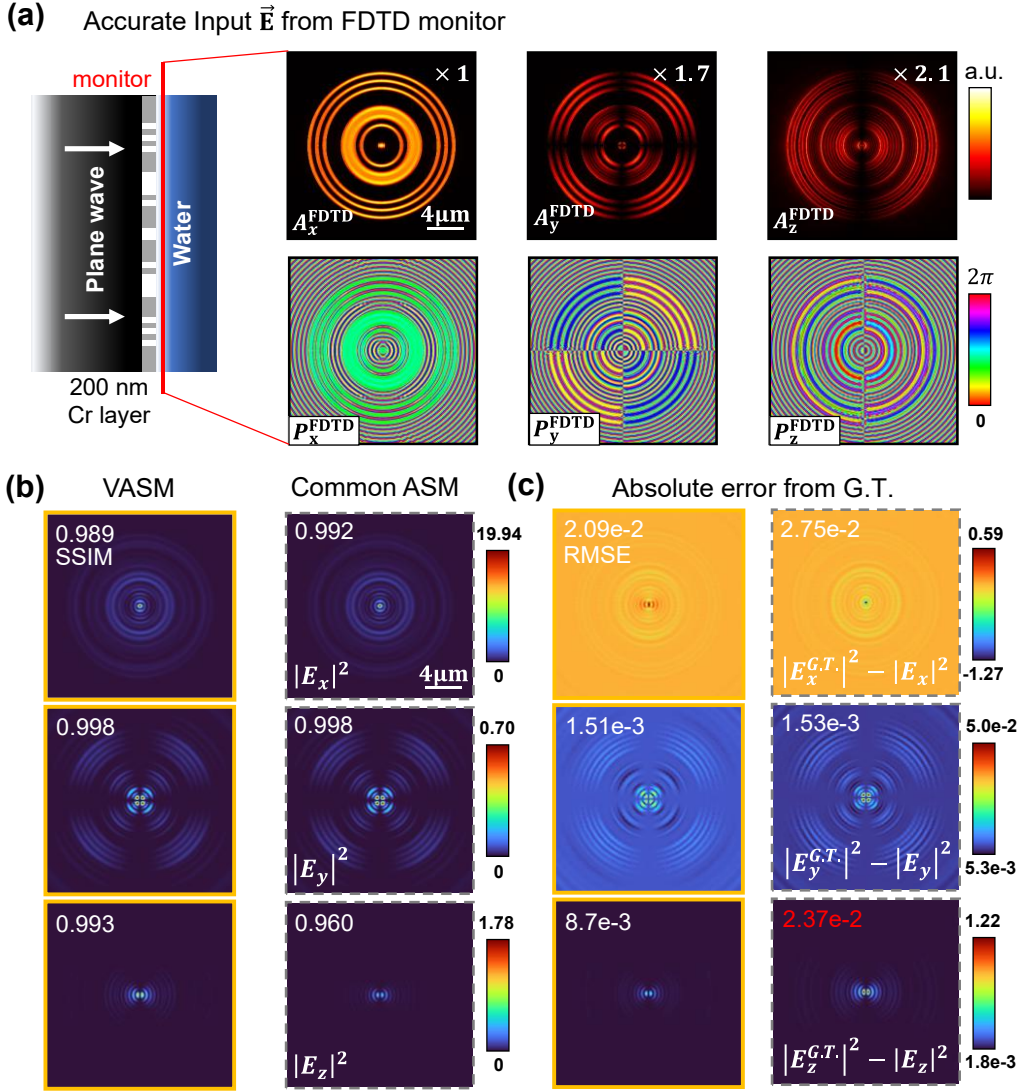

Fig. 3. A fairer comparison between VASM and common ASM. (a) More accurate input $E_x$, $E_y$, $E_z$ captured by a monitor in FDTD. First row: amplitude, abbreviated as $A$; second row: phase, abbreviated as $P$. All amplitude profiles are scaled to the same magnitude with scaling factors in the top right corner of each subplot. (b) Transversal $|\mathbf{E}|^2$ at the focal plane calculated by our VASM and common ASM, with their structure similarity index measures (SSIM) to G.T. in the top left corner of each subplot. (c) Absolute error of the $|\mathbf{E}|^2$ from G.T. (showed in Fig. 2(c)), with the total root mean square errors (RMSE) in the top left corner of each subplot.

For comparison, the results calculated by common ASM, which are framed by dashed grey boxes, failed to model the vector field. An obvious difference is that the longitudinal intensity shown in Fig. 2(c) is zero due to the absent of longitudinal input. Moreover, the absolute magnitude of the common ASM is about 1.4 times larger than our method, because the overall energy is conserved, and the excess part should have been separated into longitudinal component.

After testing the robustness of VASM, we may set a fairer comparison case for our VASM and common ASM, by replacing the ideal situation shown in Fig. 2(a) with a more accurate input near the mask directly from FDTD monitor. We put a monitor 0.1 μm after the chromium layer in FDTD to collect the absolute magnitude and phase of Cartesian field, $E_x$, $E_y$, $E_z$, as shown in Fig. 3(a), respectively. This distance was carefully chosen to balance two competing factors: maintaining sufficient proximity to capture scattering field information with a finite size while emphasizing highly interacting field components which may magnify the inaccuracies of common ASM.

Since the incident vector fields shown in Fig. 3(a) satisfy the divergence-free condition thus being well-defined, we expect that both methods should yield same results. The intensities on the focal plane calculated by VASM and common ASM are depicted in Fig. 3(b), which are framed by solid yellow boxes and dashed grey boxes respectively. As expected, both methods realize qualitative retrieval of the intensity profiles given by FDTD (G.T. shown in Fig. 2(c)), with the structure similarity index measures (SSIMs) of all three components close to 1.0, inserted in top-left corners.

However, quantitative comparisons reveal that the common ASM yield $10^0$-$10^1$ times less accuracy in longitudinal component $E_z$, while the maximum values of G.T., our approach and common one being 1.778, 1.595 and 0.557 respectively. The absolute errors distributions and the total root mean square errors (RMSE) are shown in Fig. 3(c). This unacceptable mismatch directly reveals that the common ASM fails to model the conversion from transversal components to longitudinal one, even with a well-defined incident plane.

Notably, these results demonstrate the high precision of our VASM approach, particularly in capturing for the often-neglected longitudinal component $E_z$, thereby enabling comprehensive modeling and engineering of vectorial diffractive fields in all dimensions.

*3.2 Reflection and transmission of an interface*

We further validated the effectiveness of VASM in dealing with dielectric interfaces, particularly in calculating the reflection and transmission, a remarkable challenge that can only be achieved through a closed form of projection matrix. Such calculations are quite complicated - especially in cases of high refractive index contrast or total internal reflection (TIR) where multiple interactions such as reflection, transmission, evanescent waves and optical localization effects are involved. These interactions, together with the potential numerical errors in FDTD resulting from discontinuity, bring challenges in calculating the electromagnetic fields near the interface. To make a fair and meaningful comparison, we examine the co-polarized fields 2 μm after the interface where propagating waves dominate, and localized effects are negligible.

As shown in Fig. 4(a), We examined two different media configurations: one employed a $SiO_2$ (n = 1.459)/water (n = 1.33) interface to verify the substrate effects on the diffracted field distributions shown in Fig. 3; the other employed a $SiO_2$/vacuum (n = 1.0) interface with a higher refractive index contrast to test the limits of our projection-based VASM method. The mask is located at $z = 0$. The substrate thickness $z_1$ was chosen as 2 μm, and therefore the interface is at $z = 2$ μm.

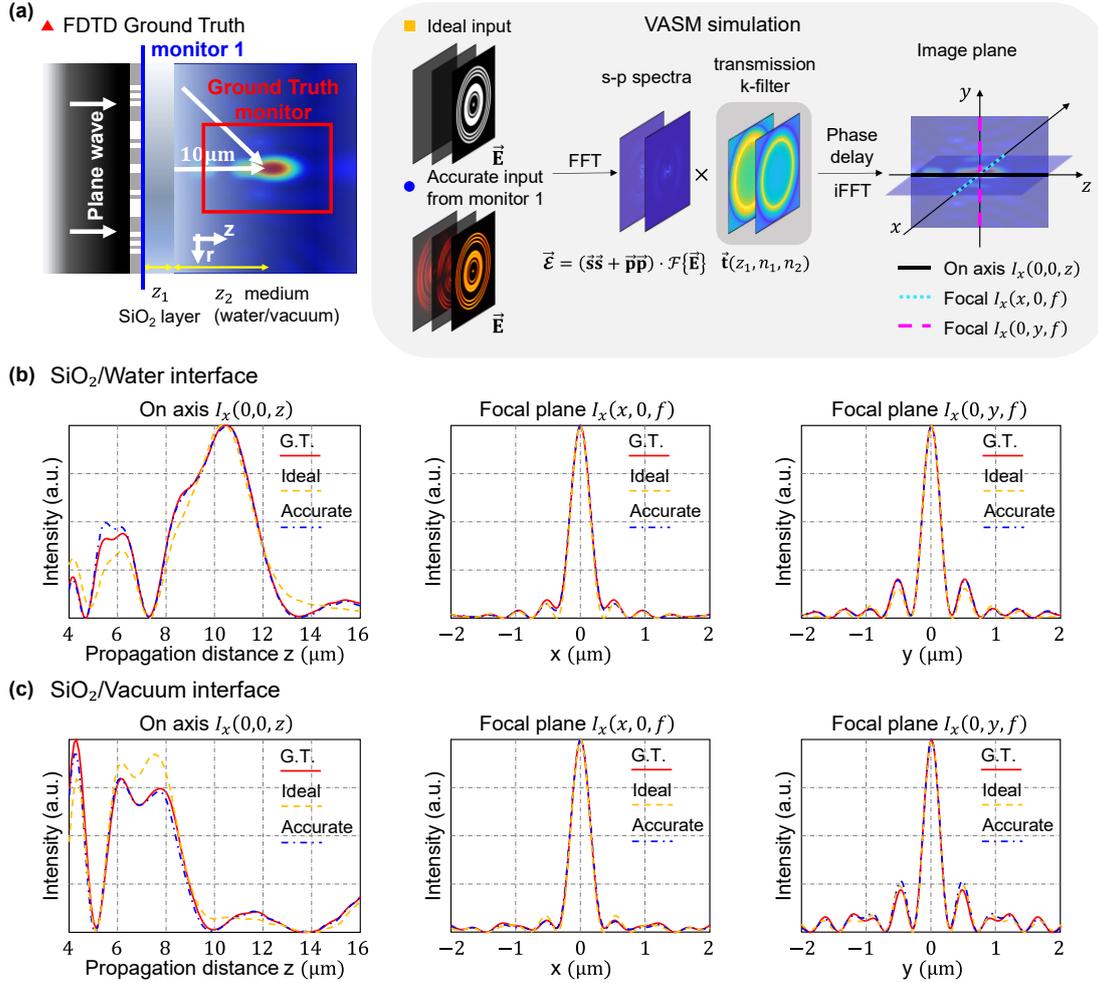

Fig. 4. A validation model of VASM with a dielectric interface. (a) Schematic of FDTD simulation and VASM simulation. An additional k-filter is added in VASM compared to free-space cases. Three representative 1D-intensity functions are chosen. (b) Focusing properties with a $SiO_2$/water interface. (c) Focusing properties with a $SiO_2$/vacuum interface.

We use full-wave FDTD simulation as ground truth (marked by a red triangle in Fig. 4(a)) and compare the results with our VASM. Simulation inputs of VASM also differs, that both ideal illumination (with only $\vec{E} = E_x \vec{x}$) and a more accurate input directly from FDTD are considered, which are marked by yellow rectangle and blue circle in Fig. 4(a), respectively. The input monitor remains 0.1 μm after the chromium layer in FDTD. As a result, the captured fields are still within the $SiO_2$ substrate, ensuring itself to undergo a filtering process in k-space (framed by dark grey box in Fig. 4(a)) and thus proving the validity of our VASM in a non-degenerated case.

We chose three representative 1D-intensity functions to quantitatively compare the dominant $I_x$ distributions: the on-axis intensity $I_x(0,0,z), z \in [4,16]$, the focal $I_x(x,0,f), x \in [-2,2]$ along x-axis and the focal $I_x(0,y,f), y \in [-2,2]$ along y-axis. The chosen ranges are plotted in Fig. 4(a) as solid black, cyan dotted and magenta dashed lines. Span of 12 μm along the z-axis and 4 μm along the x/y-axes are enough to capture the sub-foci and off-axis features.

The numerical results reveal the remarkable capabilities of VASM approach in handling dielectric interfaces. The 1D-intensity results from FDTD ground truth, VASM with ideal input and accurate input are plotted in solid red, dashed yellow and dot-dashed blue curves, respectively. In the $SiO_2$/water interface case (Fig. 4(b)), the intensity distributions show excellent agreement among all three methods, particularly in the central focal region. On-axis intensity curves show consistent peak positions and relative intensities, while the characteristic asymmetric features in transverse profiles $I_x(x, 0, f)$ and $I_x(0, y, f)$ are well reproduced, which are the hallmarks of high-NA focusing. However, further examination reveals VASM with ideal input tends to underestimate the intensity of sub-foci. It is still within our expectations since these mismatches coincide with that in free-space propagation (Fig. 2(c)), and it can be further corrected using combined method as shown.

Our VASM remains valid to model more complicated SiO$_2$/air interface case, as shown in Fig. 4(c). Despite the higher refractive index contrast and higher effective NA, the VASM maintains excellent accuracy in predicting primary focal features. Transverse intensity profiles $I_x(x, 0, f)$ and $I_x(0, y, f)$ exhibit enhanced asymmetry compared to that of Fig. 4(b), yet remain well-predicted. On the other hand, the unconformities in sub-focus peak become rather obvious. These sub-foci around 4 µm and 12 µm are highly underestimated in VASM and thus become indistinguishable. These differences exceed acceptable tolerances for high-precision applications.

Luckily, the VASM with more accurate input further corrects the deviations, showing particular performance in predicting fine field features beyond the primary focal region. The precise alignment of subsidiary focus and the preservation of relative intensity relationships demonstrate the robustness of both transmissive and free-space VASM. This achievement is especially significant, given that the method requires only a single set of input field data, making it computationally efficient compared to full FDTD simulations. Small variations in near field on-axis distributions where $z < 6$ µm (10 λ), visible in Figs. 4(b) and 4(c), provide valuable insights for future refinements while not changing the method's current practical framing.

These models validate our VASM approach for handling both free space propagation, transmission and reflection around dielectric interfaces, demonstrating accurate prediction of both primary and subsequent focal features while preserving asymmetric vectorial characteristics. The method shows robust performance across varying refractive index contrast and offers an optimal balance between accuracy and computational efficiency, making itself particularly valuable for the design and analysis of complex planar diffraction optical elements incorporating dielectric interfaces.

## 4. ROTATIONALLY SYMMETRIC PLANAR MASK

The generalized framework for calculating the diffraction field of arbitrary input is highly dependent on 2D-FFT, which would cause numerical errors and relatively long computation time when scanning along propagation axis. VASM described in Section 2 has many advantages when the diffraction mask has rotational symmetry and reduced dimension.

*4.1 Arbitrary sampling in k-space*

While 2D-FFT is mathematically rigorous, there exists numerical limits due to FFT meshing. The FFT algorithm demands an even discretization of the incident field, where the sampling steps and ranges are set in real space. The grid must be evenly partitioned along x and y axes. Consequently, the corresponding sampling in k-space is also fixed due to intrinsic reciprocity of FFT as shown in Fig. 5(a). While the size of the matrix remains consistent, the span of the k-space $\Delta k_{x,y}$ and the step in real space $dx, y$ show a reciprocal relationship: the finer sampling step in k-space, the larger span in real space, thus making large $\Delta x, y$ necessary to achieve finer sampling $dk_{x,y}$.

However, even partition in k-space leads to inaccuracy when dealing with fast oscillating phase delay propagator $\exp i\sqrt{k^2 - k_t^2}z$. Following Nyquist-Shannon sampling theorem, the sampling step in k-space must be at most a half of the oscillation period. Certainly, it would encounter an infinitesimal oscillation period at the very bandlimit $k_t = k$, thus making it impossible to capture all features near the bandlimit. Figure 5(b) illustrates the phase delay profiles $\exp i\sqrt{k^2 - k_t^2}z$ in k-space, the parameters are chosen as $\Delta x = \Delta y = 160\lambda$, $dx = dy = 0.04\lambda$, $\Delta k_x = \Delta k_y = 25\pi/\lambda$ and $dk_x = dk_y = \pi/80\lambda$. The outer black circle represents the bandlimit $\Delta k$ and the inner dashed red circle shows the highest frequency without aliasing effect, which originates from smaller oscillation period. As the propagation distance grows, the aliasing ranges also enlarges and causes more irregular profiles or periodic distortions.

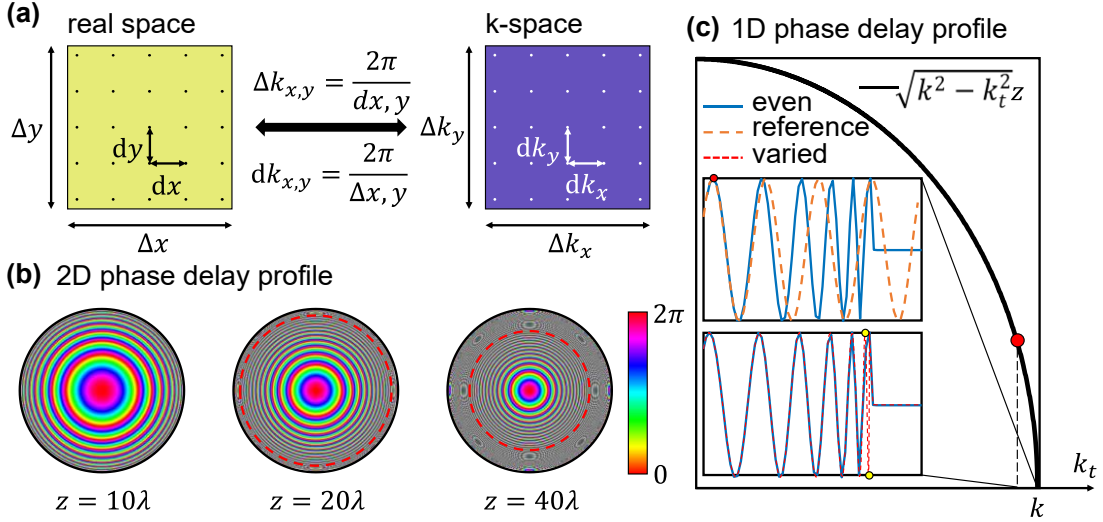

Fig. 5. FFT meshing in 2D and 1D cases. Even discretization in 2D-FFT leads to inaccuracies. (a) Meshing rules of 2D-FFT algorithm. (b) The 2D phase profiles of propagator $\exp\left(iz\sqrt{k^2 - k_t^2}\right)$. As z grows, the oscillation period becomes smaller and the area without aliasing effect shrinks. (c) The 1D phase profiles of propagator $\exp\left(iz\sqrt{k^2 - k_t^2}\right)$. Variable samplings are allowed, thus achieving extra accuracy in high frequency ranges.

While 2D-FFT requires even meshes, a rotationally symmetric mask provides a simplified 1D-Hankel transform through 1D numerical integration. Such integration allows arbitrary $dk$ and variable steps because the sampling rules in 2D-FFT are decoupled. The 1D phase delay and the sampling are plotted in Fig. 5(c), where the black curve is the actual phase delay $\sqrt{k^2 - k_t^2}z$. Higher k results in a shorter period. The period can be derived from the slope of the phase delay that $T(k_t) = |2\pi/(\partial(\sqrt{k^2 - k_t^2}z)/\partial k_t)|$, which provides a possible definition of variable step $dk(k_t) = T(k_t)/2$. Such sampling with variable sampling is plotted as red dashed curves. Oscillations in the interval $[0.95k, 1.20k]$ are plotted in inner black framed box. The dashed yellow curve shows the oscillation period of a reference red dot $k_r = 0.955k$, which fails to capture faster oscillation peaks. The blue solid curve shows the sampled peaks with an even step $dk = k/1000$. The dashed red curve shows the sampled peaks with variable step, which captures two more peaks than even sampling, indicating extra accuracy in high frequency ranges.

*4.2 Faster processing speed*

The distribution along propagation axis is conventionally acquired by repeatedly processing FFT slice by slice, which is always time-consuming. For example, we run our generalized 2D-FFT VASM code on a personal computer (CPU Inter Core i7-12700F, 16G RAM) to generate a 3D diffraction intensity distribution, with the parameters chosen as $\Delta x = \Delta y = 160\lambda$, $dx = dy = 0.04\lambda$, and $\Delta z = 20\lambda$, $dz = 0.1\lambda$. It takes about 500 seconds to collect such 4000×4000×200 3D-data library.

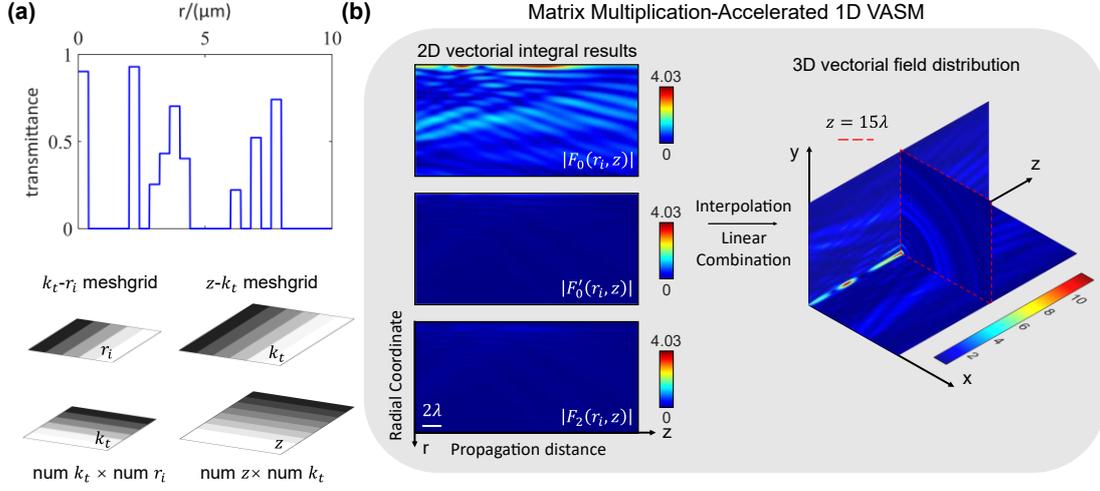

Fig. 6. The framework of generating 3D field data library of a rotationally symmetric mask through matrix multiplication-accelerated 1D integral method. (a) The required input: the annular mask distribution along radial coordinate and mesh grids. (b) The absolute values of 2D integral results and 3D vectorial field library. Only the most representative slices are visualized: the central x-z, y-z plane and the x-y focal plane at $z = 15\lambda$.

Despite the processing speed of 2D-FFT VASM has already trespassed FDTD (about 11 hours to run a 1526×1526×276 library, $\Delta x = \Delta y = 76.25\lambda, \Delta z = 13.75\lambda$, mesh accuracy 2), it can achieve even faster speed in rotationally symmetric cases. According to Eq. (9) and Eq. (11), the transversal diffracted fields of a rotationally symmetric mask can be written in three terms, the scalar part $F_0(r_i, z)$, the zeroth order vectorial part $F_0'(r_i, z)$ and the second order vectorial part $F_2(r_i, z)$.

One feasible way to accelerate the calculation is to use matrix multiplication to accelerate the integral, as shown in Fig 6. The whole procedure requires only the 1D annular mask and 2D meshing matrixes as inputs, as shown in Fig. 6(a). After proper discretization and meshing, the $F_0(r_i, z)$ term in Eq. (11) can be seen as the matrix multiplication of two parts, $J_0(k_t r_i)$ and $k^2 \widetilde{E_0}(k_t) e^{ik_z z} k_t dk_t$. The former and the latter one can be treated as a $\text{num}_{k_t} \times \text{num}_{r_i}$ and a $\text{num}_z \times \text{num}_{k_t}$ matrix, respectively. The multiplication of the two matrixes directly gives the 2D result $F_0(r_i, z)(\text{num}_z \times \text{num}_{r_i})$ of the original integration, the amplitude profiles of which are listed in Fig. 6(b) (first column) and share one same colorbar, emphasizing that the vectorial corrections only account for a small yet non-negligible part. Once these three 2D terms of integration are determined, the 3D library $\mathbf{E}(r_i, \theta_i, z)$ can be derived by linear combinations and numerical interpolations. Note that second order terms are multiplied by an exponential phase factor while processing the rotation.

This brings significant computation efficiency, and it takes only 13 seconds to generate a 2000×2000×200 3D-data library on a personal computer (CPU Inter Core i7-12700F, 16G RAM). Moreover, in matrix multiplication-accelerated 1D ASM, we can arbitrarily decide the computing range of incident plane regardless of the size of the incident plane. For example, a field data library with $\Delta x = \Delta y = 4\lambda, dx = dy = 0.01\lambda, \Delta z = 50\lambda, dz = 0.25\lambda$ (200×200×200) can be calculated within 2 (1.6~1.91) seconds.

## 5. CONCLUSION

Our study demonstrates a non-degenerate FFT-based modeling of vectorial diffraction that overcomes free divergence constraints by reconstructing a rank-deficient projection matrix. This allows for truly arbitrary incident electromagnetic fields. Compared to FDTD and common ASM, we found that traditional vector methods demand well-defined inputs, while our projection-based VASM delivers consistent results with FDTD even under simplified conditions. We also confirmed our methods' ability to treat planar interface as a k-space filter. For rotationally symmetric cases, its 2D FFT-based formulas simplify to 1D integration, achieving full 3D field libraries in seconds via matrix multiplication. Projection-based VASM's accuracy and speed make it ideal for learning-based optical designs, where existing propagation methods are slow or approximate. Notably, it models cross-polarization conversion and z-polarization behavior, enabling advanced control of vectorial fields. We anticipate our method to have a wide range of applications in fast and precise calculation of large-scale optical mask diffraction, high-NA field simulations, interface diffraction, focusing and imaging with resonant metalens, etc.

**Acknowledgment**

This work was supported by the CAS Project for Young Scientists in Basic Research (No. YSBR-049) and the Overseas Excellent Youth Science Foundation Project.

**Disclosures**

The authors declare no conflicts of interest.

**Data availability**

Codes and data underlying the results presented in this paper are available from the corresponding authors, upon reasonable request.